\documentstyle[a4,11pt]{article}

\newcommand{\sect}[1]{\setcounter{equation}{0}\section{#1}}
\newcommand{\subsect}[1]{\subsection{#1}}

\newcommand{\vs}[1]{\rule[- #1 mm]{0mm}{#1 mm}}
\newcommand{\hs}[1]{\hspace{#1 mm}}

\newcommand{\lbl}[1]{\label{eq:#1}}
\newcommand{\rf}[1]{(\ref{eq:#1})}
\newcommand{\nn}{\nonumber}

\newcommand{\be}{\vs{2}\large\begin{equation}}
\newcommand{\ee}{\vs{2}\end{equation}\normalsize}

\newcommand{\bea}{\large\begin{eqnarray}}
\newcommand{\ena}{\end{eqnarray}\normalsize}
\newcommand{\nnbea}{\large\begin{eqnarray*}}
\newcommand{\nnena}{\end{eqnarray*}\normalsize}

\newcommand{\lra}{\ \longrightarrow\ }
\newcommand{\Lra}{\ \Longrightarrow\ }
\newcommand{\Llra}{\ \Longleftrightarrow\ }
\newcommand{\ovl}[1]{\overline{#1}}

\newcommand{\mbf}[1]{\mbox{\boldmath $#1$}}
\newcommand{\tens}{\! \otimes \!}

\newcommand{\shalf}{\textstyle{\frac{1}{2}}\displaystyle}
\newcommand{\dps}{\displaystyle}
\newcommand{\nms}{\normalsize}
\newcommand{\cm}{\hspace{1cm}}

\newcommand{\bz}{{\bar{z}}}

\newcommand{\del}{\partial}

\newcommand{\zer}[1]{\stackrel{\circ}{#1}}
\def\ie{{\em i.e. }}
\def\eg{{\em e.g. }}
\def\etal{{\em et al. }}
\newtheorem{rem}{Remark}[section]
\newtheorem{defi}[rem]{Definition}
\newtheorem{thm}[rem]{Theorem}

\begin{document}

\font\fifteen=cmbx10 at 15pt
\font\twelve=cmbx10 at 12pt

\begin{titlepage}

\begin{center}

\renewcommand{\thefootnote}{\fnsymbol{footnote}}

{\twelve Centre de Physique Th\'eorique\footnote{
Unit\'e Propre de Recherche 7061
}, CNRS Luminy, Case 907}

{\twelve F-13288 Marseille -- Cedex 9}

\vspace{1 cm}

{\fifteen FLAT COMPLEX VECTOR BUNDLES, \\
THE BELTRAMI DIFFERENTIAL AND {\LARGE $W$}--ALGEBRAS}
\footnote{Work dedicated to the Memory of my friend and physicist
Tanguy ALTHERR, February 1963--July 14, 1994.}
\footnote{Work partly supported by the Commission of the European Communities
under contract Nr.ERBCHBICT930301.}

\vspace{0.3 cm}

\setcounter{footnote}{0}
\renewcommand{\thefootnote}{\arabic{footnote}}

{\bf 
Serge LAZZARINI
\footnote{and Universit\'e de la M\'editerran\'ee, Aix-Marseille II.}
\footnote{e-mail : {\tt sel@cpt.univ-mrs.fr}}
}
\end{center}

\indent

\centerline{\bf Abstract}

Since the appearance of the paper by Bilal \etal in
 '91, \cite{BFK91}, it has been widely assumed that
$W$--algebras originating from the Hamiltonian reduction of an
$SL(n,\mbf{C})$-bundle over a Riemann surface give rise to a flat
connection, in which the Beltrami differential may be identified.

In this letter, it is shown that the use of the Beltrami parametrisation 
of complex structures on a compact Riemann surface over which flat complex 
vector bundles are considered, allows to construct the above mentioned
flat connection. It is stressed that the modulus of the Beltrami 
differential is necessarily less than one, and that solutions of the so-called
Beltrami equation give rise to an orientation preserving smooth
change of local complex coordinates.
In particular, the latter yields a smooth
equivalence between flat complex vector bundles.
The role of smooth diffeomorphisms which induce 
equivalent complex structures is specially emphasized.

Furthermore, it is shown that, while the construction given here
applies to the special case of the Virasoro algebra, 
the extension to flat complex vector bundles of arbitrary rank does
not provide ``generalizations'' of the Beltrami differential usually
considered as central objects for such non-linear symmetries. 

\indent

\noindent 1991 AMS Classification: 32G07,32G08, 32G81, 32L10, 53B10, 53B50

\noindent 1995 PACS Classification: 02.40.Dr, 02.40.Hw, 11.15.-q, 11.25.Hf

\indent


\indent

\noindent February 1996, corrected version January 1997. 

\noindent CPT-96/P.3311, {\em Published in Lett. Math. Phys. {\bf 41:}
207-225, 1997.}

\indent

\noindent anonymous ftp or gopher: cpt.univ-mrs.fr

\end{titlepage}
\setcounter{footnote}{0}

\sect{Introduction and Motivation}

\indent

In a paper by A.M. Polyakov \cite{Pol90}, diffeomorphism transformations 
for the 2-d conformal geometry are identified by a soldering
procedure from 
$SL(2)$ gauge transformations. Several further works, trying to 
interpret and generalize this ad hoc process, rely on the
so-called Hamiltonian reduction. Among these papers,
that by A. Bilal and coworkers in 1991, \cite{BFK91}, 
pioneered the attempt to give
a geometric interpretation to the non-linear $W$--symmetries. 
However, in spite of the abundance of new ideas and concepts
presented in \cite{BFK91}, it seems that the first part of the paper
in fact concerns two disconnected topics. 

On the one hand, in a self-contained introduction, the authors discuss
both the notion of complex structures parametrized by Beltrami
differentials, \ie a $(-1,1)$-differential $\mu$ submitted to the
condition $|\mu|<1$, and the reinterpretation of the compatibility
condition between complex and projective structures as the
vanishing of the curvature 2-form of a connection. This connection has
components which depend explicitly on $\mu$.

On the other hand, in the part concerned with the Hamiltonian
reduction, a general gauge scheme is provided as a possible way of
generating $W$--algebras. But, when restricted to the Virasoro case,
even if the above quoted connection is formally recovered, the
condition $|\mu|<1$ is not. This seems to be a technical detail.
However, one ought to have expected that,
especially in the simplest case of the Virasoro algebra, the requirement
$|\mu|<1$ should have been recovered. This lack of consistency
is also encountered in other papers on the subject, see \eg
\cite{Ver90,OSSvN92,dBG93} and references therein. In these articles, the
identification of an ``algebraically'' obtained quantity carrying the
same geometrical properties as a Beltrami differential is often made
without paying any attention to the extra condition $|\mu|<1$.
If this observation that there are ``Beltrami-like'' objects is correct, 
then, clearly, it questions the geometric framework of
\cite{BFK91} from which $W$--algebras are assumed to originate.

This question is important because the Beltrami differential
is one of the central objects in the Lagrangian description of 2-d
conformal models, \cite{BB87,Bec88,Sto88,KLS91}. 
The requirement that $|\mu|<1$ is
also related to the positivity of the conformal
classes of metrics on the surface, $ds^2\propto |dz + \mu\,d\ovl{z}|^2$,
\cite{Leh87}, and the fact that projective structures are parametrized by 
holomorphic projective connections \cite{Gun66,Gun67b}. Following 
\cite{BFK91}, the latter are readily identified with the
well-known spin two ``energy-momentum tensor'' of bidimensional
conformal models. Generalization of that concept to higher spin objects
initiated by A.B.~Zamolodchikov, \cite{Zam85}, yields the so-called
non-linear $W$--algebras as extensions of the 2-d conformal algebra,
\ie the Virasoro algebra.

In this letter, it is shown that there is a consistent geometrical
construction in which both the gauge aspect and the
desired property of the Beltrami differential are reconciled.
Two steps are required for a proper treatment of the geometry. 
In the first step, the gauge scheme approach 
of Bilal \etal is reconsidered in the
theory of flat complex vector bundles over a Riemann surface. In this
holomorphic framework, as will be seen, special representatives for holomorphic
connections turn out to have the so-called Drinfeld-Sokolov form \cite{DS84}. 
In the second step, we give up the hamiltonian reduction method proposed
in \cite{BFK91}, in favour of the study of a smooth change of local
complex coordinates which implements the Beltrami
parametrisation of complex structures. When the latter is used, the
action of smooth diffeomorphisms on the Riemann surface is well
manageable.
It is recalled that the diffeomorphism group is the symmetry group for 
the 2-d Lagrangian conformal models whose extensions are expected to be
$W$--algebras. 

The study of holomorphic vector bundles has been carried out
intensively in the context of $W$--algebras 
in \cite{Zuc94a,Zuc94b} where the role of the smooth change of complex
coordinates has mostly been overlooked. The flat complex 
vector bundles involved turn out to be jet bundles of the sheaf of 
solutions of some conformally covariant complex differential equations. 
These results will make contact with some of those obtained in \cite{BFK91}.
Nevertheless, we shall restrict ourselves to the geometric aspects of the
construction without speculating on the possible physical interpretation
of the quantities involved, even if this deserves some careful attention.

In the case of the Virasoro algebra, the construction works perfectly.
However, its generalization to the $W$--case unfortunately
does not lead to objects extending the notion of the Beltrami
differential \cite{BFK91,AG92,Zuc93b,GGL95}, even if the latter
is truly involved, as will be shown in Section 5. 
This suggests that the geometry of $W$--algebras, if any,
is far from being understood. To be precise, there is a clash
between, on the one hand, the desire to retain the Beltrami
parametrization of complex structures which, as already said, is crucial
for the study of bidimensional conformal models, and, on the other hand, the
desire to extend the conformal symmetry to the $W$--algebras.  

The letter is organized as follows.
In section two, we shall describe the basic material and general
results of Gunning's work on flat complex vector
bundles. This deals with the gauge aspects of the construction.
Section three is devoted to the equivalence between complex structures
within the Beltrami parametrization with special emphasis that the
latter will implement a local smooth change of complex coordinates
ont the Riemann surface. Infinitesimal features 
of the action of diffeomorphims will be briefly analyzed in the BRS framework.
Section four is concerned with $SL(2,\mbf{C})$ complex vector bundles
which happen to reproduce, when the action of diffeomorphisms is considered,
exactly the expected results for the case of the Virasoro algebra. 
In the fifth Section, the question of extending the use of flat
complex vector bundles is addressed. The study of the $SL(3,\mbf{C})$ 
complex vector bundles is carried out, with the Beltrami differential 
still involved. The description of the $W_3$--case is only partially
obtained. The letter concludes with a few words regarding
the geometric content of $W$--algebras.

\indent

\sect{A short review of Gunning's results on flat complex\\ vector bundles}

\indent

We will use Gunning's notation in order to make contact with his
results \cite{Gun66,Gun67b,Gun67a}. Gunning's results hold in the
holomorphic case, \ie both gauge group and connection will be holomorphic.

Consider a compact Riemann surface $\Sigma$ of genus greater than one
with a complex coordinate covering $\{(U_{\alpha},Z_{\alpha})\}$.
Throughout this work, capital letters will label the geometry for the
complex structure corresponding to the local complex analytic
coordinates $\{Z_{\alpha}\}$. Let $K$ be the canonical bundle of
$\Sigma$ defined by the holomorphic 1-cocycle $K_{\alpha\beta} =
dZ_{\beta}/dZ_{\alpha}$. We furthermore consider over $\Sigma$ the
holomorphic vector bundle $\Phi$ of rank $n$ determined by holomorphic
transition functions $\Phi_{\alpha\beta}$ (1-cocycle with respect to
the above covering). According to Gunning \cite{Gun67a,Gun67b}, we define

\begin{defi}
The endomorphisms of $\Phi$ are defined by the following set
$\{G_{\alpha}\}$ of matrix-valued functions glueing as
\be
\mbox{\nms in } U_{\alpha} \cap U_{\beta}\ :\
G_{\alpha} \Phi_{\alpha\beta} \ =\ \Phi_{\alpha\beta}G_{\beta} \ .
\lbl{aut}
\ee
\end{defi}
At this stage we still have the freedom to choose either a smooth or a
holomorphic gauge group according to the choice of smooth or
holomorphic sections of the bundle $\Phi$.

\begin{defi}
A connection $A$ on the vector bundle $\Phi$ is a collection of
matrix-valued differential 1-forms, $\{A_{\alpha}\}$, defined by the
following patching rules
\be
\mbox{\nms in } U_{\alpha} \cap U_{\beta}\ :\
d \Phi_{\alpha\beta} = \Phi_{\alpha\beta} A_{\beta} 		-
A_{\alpha} \Phi_{\alpha\beta} \ .
\lbl{conn}
\ee
\end{defi}
Since the gauge group acts on the fibres of $\Phi$ it will also acts
on connection by
\begin{defi}
The gauge transformed of $A$ is locally defined by
\be
\mbox{\nms in } U_{\alpha}\ :\
({}^G A)_{\alpha}\ =\ G_{\alpha} A_{\alpha} G_{\alpha}^{-1} +
G_{\alpha} d G^{-1}_{\alpha} \ .
\lbl{gaugetrsf}
\ee
\end{defi}

First of all, the structure of connections can be analysed. The
$(0,1)$-component $\ovl{A}$ of any connection on a complex vector
bundle over a Riemann surface is integrable so that there is a local
change of the fibre coordinates for which the $(0,1)$-component may be
put to zero \cite{Ati57,AB82}.

So, we have the following
\begin{thm}
At fixed complex structure on $\Sigma$, the compatible complex
analytic structures on the bundle $\Phi$ viewed as a smooth vector
bundle, are in one-to-one correspondence with the gauge equivalence
classes of connections of type $(0,1)$.
\end{thm}
Integrating the $(0,1)$-component of the connection means finding a
collection of matrix-valued functions $\{M_{\alpha}\}$, defined up to
an holomorphic matrix-valued function, such that locally
\be
\mbox{\nms in } U_{\alpha}\ :\
\ovl{A}_{\alpha}\ =\ M^{-1}_{\alpha} \del_{\ovl{Z}_{\alpha}} M_{\alpha} \ .
\ee
One obtains a {\em new} complex vector bundle equivalent to $\Phi$
and defined by the 1-cocycle $M_{\alpha} \Phi_{\alpha\beta}
M^{-1}_{\beta}$. The latter defines new coordinates for which 
the $(0,1)$-component of any connection vanishes according to the
local redefinition
\be
\mbox{\nms in } U_{\alpha}\ :\
{\cal A}_{\alpha}\ =\ (M_{\alpha} A_{\alpha} - dM_{\alpha})M_{\alpha}^{-1}\ .
\lbl{redef}
\ee
This redefinition should not be confused with a gauge transformation 
\rf{gaugetrsf}. One can show that $M_{\alpha} \Phi_{\alpha\beta} =
N_{\beta}M_{\beta}$ in the overlap $U_{\alpha} \cap U_{\beta}$, where 
$N_{\beta}$ is a holomorphic matrix-valued function, and thus means that
the local solutions $\{M_{\alpha}\}$ are no longer endomorphisms of $\Phi$.
This new complex structure does not depend on the
choice of the representative for $\ovl{A}$.

From now on, the coordinate system on the holomorphic vector bundle,
with {\em holomorphic} gauge group, will be so chosen
that $\ovl{A}\equiv 0$, \ie a connection on $\Phi$ is then a
collection of matrix-valued differential forms of type $(1,0)$ glueing
according to \rf{conn}. Note that the vanishing of the $(0,1)$ part of
\rf{conn} makes sense thanks to the holomorphicity of $\Phi_{\alpha\beta}$. 

A {\em flat} bundle associated to the complex bundle $\Phi$ will be a
complex analytic bundle equivalent to $\Phi$ with constant transition
functions. Flat connections thus become holomorphic in those holomorphic
coordinates. 

\begin{thm}[Gunning, \cite{Gun67a}]
The set of flat vector bundles associated to a complex vector bundle
$\Phi$ is in one-to-one correspondence with the gauge equivalent
classes of holomorphic $(1,0)$-connections. Moreover if $\det\Phi=1$,
the corresponding set of flat vector bundles is in one-to-one
correspondence with the gauge equivalent classes of holomorphic
$(1,0)$-connections with null trace.
\end{thm}
{\bf N.B.} Since $\ovl{A}\equiv 0$, one has $F(A) =0 \Llra
\ovl{\del}A=0$. If furthermore $\det\Phi=1$ then $\Phi$ is determined by
an $SL(n,\mbf{C})$ 1-cocycle.

This sums up the main results concerned with flat holomorphic vector
bundles, \ie the study of holomorphic $(1,0)$-connections.
Moreover, as a nice feature, there are always representatives of
$\Phi$ such that the matrices $\Phi_{\alpha\beta}$ and $G_{\alpha}$
are all upper triangular \cite{Gun67b}. So, in the examples
thereafter treated, we can limit ourselves with this special upper
triangular form.

\indent

\sect{Equivalence of complex structures}

\indent

In order to speak about the equivalence of complex structures, it is 
rather good to introduce the Beltrami parametrisation of complex
structures, see \eg \cite{Leh87}, subordinate to the smooth structure 
of the surface $\Sigma$. 
It turns out that some complex structures can be defined to be equivalent in
a sense which will be made precis down below.

In doing so,
we need another diffeomorphic copy of $\Sigma$ according to the unique
differential structure on $\Sigma$. This copy is turned into a 
Riemann surface, $\Sigma_0$, by choosing a fixed background complex 
structure given by the complex covering $\{(U_{\alpha},z_{\alpha})\}$. 
We will denote by $\kappa$ the canonical bundle of $\Sigma$ defined by
the 1-cocycle $\kappa_{\alpha\beta} = dz_{\beta}/dz_{\alpha}$ with 
respect to this fixed local complex coordinates $\{z_{\alpha}\}$, 
and we will set $\del_{\alpha} \equiv \del/\del z_{\alpha}$ and 
$\ovl{\del}_{\alpha} \equiv \del/\del \ovl{z}_{\alpha}$. 

We introduce a Beltrami differential denoted by $\mu$, namely a 
$(1,0)$-vector valued $(0,1)$-form with $|\mu|<1$. Strickly speaking, 
the Beltrami differential $\mu$ can be seen as a smooth section of the bundle
$\kappa^{-1}\tens\ovl{\kappa}$
\be
\mbox{\nms in } U_{\alpha} \cap U_{\beta}\ :\
\mu_{\alpha}\ =\ 
\kappa_{\alpha\beta}^{-1}\, \ovl{\kappa}_{\alpha\beta}\, \mu_{\beta}\ , 
\ee 
such that locally $|\mu_{\alpha}|<1$. Let ${\cal B}(\Sigma)$ denote
the space of smooth Beltrami differentials on $\Sigma$.

\subsect{The smooth change of local complex coordinates}

\indent

The Beltrami parametrization of complex structures over $\Sigma$, consits in
finding the holomorphic coordinates $\{Z_{\alpha}\}$ pertaining to the
complex structure parametrized by $\mu\equiv\{\mu_{\alpha}\}$,
$|\mu_{\alpha}|<1$. This amounts to
solve locally the following Pfaff system, see \eg \cite{Sto88},
\be
\mbox{\nms in } U_{\alpha}\ :\
dZ_{\alpha}\ =\ \lambda_{\alpha} (dz_{\alpha} + \mu_{\alpha}
d\bz_{\alpha})
\cm \Lra \del_{Z_{\alpha}} = 
\frac{\del_{\alpha} - \ovl{\mu}_{\alpha}
\ovl{\del}_{\alpha}}{\lambda_{\alpha}(1 - \mu_{\alpha}\ovl{\mu}_{\alpha})} \ ,
\lbl{dZ}
\ee
with integrating factor $\lambda_{\alpha}=\del_{\alpha}Z_{\alpha}$
fulfilling
\be
\mbox{\nms in } U_{\alpha}\ :\
(d^2 = 0 \Llra )\cm (\ovl{\del}_{\alpha} - \mu_{\alpha}
\del_{\alpha})\ln\lambda_{\alpha} = \del_{\alpha}\mu_{\alpha} \ .
\lbl{lamb}
\ee
Solving the Pfaff system \rf{dZ} is equivalent to solving locally 
the so-called Beltrami equation
\be
\mbox{\nms in } U_{\alpha}\ :\
(\ovl{\del}_{\alpha} - \mu_{\alpha}\del_{\alpha})\, Z_{\alpha}\ =\ 0\ .
\lbl{beltra}
\ee
According to Bers, see \eg \cite{Leh87}, the Beltrami equation
\rf{beltra} always admits as a solution a quasiconformal mapping
with dilatation coefficient $\mu_{\alpha}$. One thus remarks that $Z_{\alpha}$
is a holomorphic functional of $\mu_{\alpha}$, which will be denoted
for a while by $Z_{\mu_{\alpha}}$,
and will play the role of the complex coordinate
introduced in the previous Section. Therefore, the solution of the
Beltrami equation is a mapping on
$U_{\alpha}\ :\ (z_{\alpha},\bz_{\alpha}) \lra
(Z_{\alpha}(z_{\alpha},\bz_{\alpha}),
\ovl{Z}_{\alpha}(z_{\alpha},\bz_{\alpha}))$ 
which preserves the orientation
(the latter condition secures the requirement $|\mu_{\alpha}|<1$ as 
noticed in \cite{Zuc93b}), and is thus locally invertible, so that
$Z_{\mu_{\alpha}}(z_{\alpha},\bz_{\alpha})$ defines a new complex coordinate
mapping on the open set $U_{\alpha}$. Moreover, in the intersection
$U_{\alpha} \cap U_{\beta}$, it follows from the patching law of
$\mu$, that solving the Beltrami equation in the overlap shows
that the transition function $Z_{\mu_{\alpha}}\circ Z_{\mu_{\beta}}^{-1}$
is holomorphic in $Z_{\mu_{\beta}}$ and depends holomorphically on
$\mu$. Hence, the covering 
$\{(U_{\alpha},Z_{\mu_{\alpha}}\})$ defines a new complex structure on
$\Sigma$, the one given by $\mu$ and denoted by $\Sigma_\mu$. 
The fibered complex manifold ${\cal B}(\Sigma) \times \Sigma$, with local
complex coordinates $(\mu,Z_\mu)$, defines a complex analytic family
of compact Riemann surfaces, see Chapter 2 of \cite{Kod86}. 
The reader is also referred for instance to Appendix D of \cite{STW95}.
We will say that the Riemann surface $\Sigma_{\mu}$ is different from
the Riemann surface $\Sigma_0$, ($\mu \equiv 0$ corresponds
to the standard complex structure), with local complex
coordinates $\{z_{\alpha}\}$. The appearance of the Beltrami
differential simply stems from this  non algebraic (!) process 
implemented by this (local) smooth change of local complex coordinates.

\subsect{The action of smooth diffeomorphisms}

\indent

One remark is in order. The Beltrami differential is a geometric
object and as such, is transformed under the action of smooth 
diffeomorphisms by pull-back. Let $\varphi$ be a smooth 
diffeomorphism of $\Sigma$ homotopic to the identity map.
Let $\mu$ be a given smooth Beltrami differential on $\Sigma$,
represented in a chart, let say $(U_{\alpha},z_{\alpha})$, by the
smooth function $\mu_{\alpha}$.
Local complex coordinates on the inverse image of $U_{\alpha}$ 
by $\varphi$, $\varphi^{-1}(U_{\alpha})$, are provided
by intersection with a chart, $(\varphi^{-1}(U_{\alpha})\cap
U_{\beta},z_{\beta})$. According to the fixed complex structure
on $\Sigma$, the local representative of $\varphi$ reads
\be
(\varphi^z_{\alpha\beta}(z_{\beta},\bz_{\beta}),
\varphi^{\bz}_{\alpha\beta}(z_{\beta},\bz_{\beta}))
\ =\ (z_{\alpha} \circ \varphi,\bz_{\alpha}\circ \varphi )\ .
\ee
The pull-back, $\mu^{\varphi}$, of $\mu$ by $\varphi$ is a new section
of $\kappa^{-1}\tens\ovl{\kappa}$, with $|\mu^{\varphi}|<1$ since 
$\varphi$ is homotopic to the identity, see for instance \cite{Nag88}. 
In terms of components and coordinates, $\mu^{\varphi}$ is given on 
$\varphi^{-1}(U_{\alpha})\cap U_{\beta}$ by 
\be
(\mu^{\varphi})_{\beta}\ =\ 
\frac{ \ovl{\del}_{\beta} \varphi^z_{\alpha\beta}  
+ \ovl{\del}_{\beta} \varphi^{\bz}_{\alpha\beta}\,
(\mu_{\alpha} \circ \varphi)}
{\del_{\beta} \varphi^z_{\alpha\beta} 
+ \del_{\beta} \varphi^{\bz}_{\alpha\beta}\,
(\mu_{\alpha} \circ \varphi)}\ .
\lbl{diffmu}
\ee
We get for any point in $U_{\alpha}$ the following commutative diagram
\be
\begin{array}{ccc}
(\varphi^z_{\alpha\beta}(z_{\beta},\bz_{\beta}),
\varphi^{\bz}_{\alpha\beta}(z_{\beta},\bz_{\beta})) 
& \stackrel{\mu_{\alpha}}{-\hs{-0.8}-\hs{-2}\longrightarrow} 
& (Z_{\mu_{\alpha}}\circ \varphi,\ovl{Z}_{\mu_{\alpha}}\circ \varphi) 
\\[2mm]
\uparrow & & \uparrow \\[-2mm]
\mid & & \mid \\[-2mm]
\mid & & \mid \\[2mm]
(z_{\beta},\bz_{\beta}) & 
\stackrel{(\mu^{\varphi})_{\beta}}{-\hs{-0.8}-\hs{-2}\longrightarrow} 
& (Z_{(\mu^{\varphi})_{\beta}}(z_{\beta},\bz_{\beta}),
\ovl{Z}_{(\mu^{\varphi})_{\beta}}(z_{\beta},\bz_{\beta}))
\end{array}
\lbl{diag}
\ee
by solving the Beltrami equation for both $\mu$ in $U_{\alpha}$
and $\mu^{\varphi}$ in $\varphi^{-1}(U_{\alpha})\cap U_{\beta}$.
Since the action of diffeomorphisms on Beltrami differentials is
defined by the pull-back \rf{diffmu}, it can be shown \cite{Leh87}
that the right vertical arrow is a
biholomorphic mapping and hence $Z_{(\mu^{\varphi})_{\beta}}$ 
can be chosen in $\varphi^{-1}(U_{\alpha})\cap U_{\beta}$ as 
$Z_{(\mu^{\varphi})_{\beta}} = Z_{\mu_{\alpha}}\circ \varphi$.
One can see that diffeomorphisms act holomorphically on the
coordinates $(\mu,Z_\mu)$, and therefore $\mu$ and $\mu^{\varphi}$
define equivalent complex structures.

As will be seen later on by explicit computation, for a given complex 
structure $\mu$, the action of diffeomorphisms on geometric objects
related to the holomorphic coordinates $Z_{\mu}$, 
(germs of holomorphic sections, abelian or holomorphic quadratic
differentials, etc, and generically denoted here by $\Pi$) is easily worked
out by pull-back
\be
(\varphi^*(\Pi))_{Z_{\mu^{\varphi}}} (Z_{\mu^{\varphi}})\ =\
\Pi_{Z_{\mu}} (Z_{\mu}\circ\varphi)\ .
\lbl{geobj}
\ee
It is also worthwhile to consider the infinitesimal action of
diffeomorphisms. 

\subsect{The infinitesimal action of diffeomorphisms}

\indent

The infinitesimal action of diffeomorphisms on $\Sigma$ reduces down to
representing the Lie algebra of diffeomorphisms connected to the identity
and thus, preserving the orientation. The BRS formulation 
in terms of a nilpotent algebraic operation $s$ will be used for the
(space) variations given by the Lie derivative, (with respect to the
background complex coordinates $\{z_{\alpha}\}$), of geometric objects
which will be involved in the exercise given below.

In this setting, the infinitesimal action of diffeomorphisms on a
Beltrami differential $\mu$ is now well-known \cite{BB87,Bec88}. But,
another way to derive the variation which will be quite useful for
our purpose can be directly obtained from the above diagram \rf{diag}.

In the BRS algebraic set-up, we have $s d + d s = 0$, where $d$ is the
exterior derivative. According to the complex coordinate covering
$\{(U_{\alpha},Z_{\alpha})\}$ pertaining to the complex structure
given by $\mu$ where ``everything'' will be taken to be holomorphic, one
considers the action of smooth diffeomorphisms homotopic to the
identity map, $\varphi_t = id_{\Sigma} + t\, c + o(t)$, where 
$c = c^z\,\del + c^{\ovl{z}}\,\ovl{\del}$ is the smooth Faddeev-Popov
ghost associated to vector fields\footnote{Mathematically speaking, 
$c$ is the generator of the Grassmann algebra of the dual of the Lie 
algebra of smooth diffeomorphisms of $\Sigma$.}
with respect to the background complex coordinates $\{z_{\alpha}\}$, 
$s\,z_{\alpha}=0$. As said before, the action produces a new equivalent
complex structure to that of $\mu$ given by $\mu^{\varphi_t}$ and for 
which the complex covering can be written as
$\{(\varphi^{-1}(U_{\alpha}),Z_{\mu_{\alpha}}\circ \varphi_t)\}$, so that
$Z_{\mu}$ is transformed as $Z_{\mu}\circ\varphi_t = Z_{\mu^{\varphi_t}}$.

The infinitesimal action given by the (graded) Lie derivative,
$s \equiv L_c = i_c\, d - d\, i_c$ and, thanks to \rf{dZ}, 
writes locally in $U_{\alpha}$, 
\be
sZ_{\alpha}\ =\ L_c Z_{\alpha}\ =\ i_c d\, Z_{\alpha}\ 
=\ \lambda_{\alpha} (c^{z_{\alpha}}+\mu_{\alpha}\,c^{\ovl{z}_{\alpha}}) 
\equiv \lambda_{\alpha}\,C_{\alpha} \equiv \gamma^{Z_{\alpha}}\ ,
\ee
where $c = \gamma^Z \del_Z + \gamma^{\ovl{Z}} \del_{\ovl{Z}}$ is
the (smooth) Faddeev-Popov ghost in the holomorphic coordinates $Z$,
and $C_{\alpha}$ reflects a change of basis 
in the Lie algebra of diffeomorphisms leading to the holomorphic factorization
property of bidimensional conformal models \cite{BB87,Bec88,KLS91}.

So, with respect to the complex structure $\mu$, the infinitesimal action of
diffeomorphisms writes locally in $U_{\alpha}$, with $s^2=0$,
\bea
&&s Z_{\alpha} \equiv \gamma\cdot Z_{\alpha}\ =\ \gamma^{Z_{\alpha}},\cm
s \gamma^{Z_{\alpha}}\ =\ 0,\nn\\[2mm]
&& s \,d+d\,s=0 \Lra
[s,\del_{ Z_{\alpha}}] = 
- (\del_{Z_{\alpha}}\gamma^{Z_{\alpha}})\del_{Z_{\alpha}}
- (\del_{Z_{\alpha}}\gamma^{\ovl{Z}_{\alpha}})\del_{\ovl{Z}_{\alpha}}\ .
\lbl{sZ}
\ena
Furthermore, the infinitesimal version of the action of diffeomorphisms on 
(holomorphic) geometric objects defined by \rf{geobj}, is given by the
Lie derivative which takes in the holomorphic coordinates $Z$, 
the following very simple expression 
\be
s\, \Pi_{\alpha}\ =\ (L_{c}\Pi)_{\alpha}\ =\ 
\gamma^{Z_{\alpha}} \del_{Z_{\alpha}} \Pi_{\alpha}\ .
\lbl{infgeobj}
\ee
Next, by computing 
\be
s\, d\, Z_{\alpha}\ =\ - d \gamma^{Z_{\alpha}}\ =\ 
s ( \lambda_{\alpha} (dz_{\alpha} + \mu_{\alpha} d\bz_{\alpha}))\ ,
\ee
see \rf{dZ}, together with the compatibility condition
\rf{lamb}, gives rise to the following infinitesimal action of 
diffeomorphisms in terms of the nilpotent BRS operation, $s^2 = 0$, 
\be
s \mu_{\alpha}\ =\ (\ovl{\del}_{\alpha} - \mu_{\alpha} \del_{\alpha}
+ \del_{\alpha}\mu_{\alpha}) C_{\alpha}\ ,\ \
s C_{\alpha}\ =\ C_{\alpha} \del_{\alpha} C_{\alpha}\ ,
\ \ s\ln\lambda_{\alpha} = \del_{\alpha}C_{\alpha} 
+ C_{\alpha}\del_{\alpha}\ln\lambda_{\alpha}\ .
\lbl{sz}
\ee
Remark that, through the definition of $C_{\alpha}$,
the infinitesimal variation of $\mu_{\alpha}$ given in \rf{sz}
equals the infinitesimal action of diffeomorphisms directly computed
from the formula \rf{diffmu}.

At this stage, we have settled all the basic ingredients we need to
deal with the Virasoro algebra.

\indent

\sect{The $SL(2,\mbf{C})$ complex vector bundle}

\indent

To start with the usual Virasoro case, one considers over $\Sigma$,
according to Section two, the holomorphic vector bundle $\Phi$ 
of rank 2 defined by the following holomorphic transition functions
\be
\mbox{\nms in } U_{\alpha} \cap U_{\beta}\ :\
\Phi_{\alpha\beta}\ =\ \left(
\begin{array}{cc}
K_{\alpha\beta}^{1/2} &
{\dps \frac{dK_{\alpha\beta}^{1/2}}{dZ_{\beta}}}\\[4mm]
0 & K_{\alpha\beta}^{-1/2}
\end{array}
\right)\ .
\ee
Note that $\det \Phi = 1$.

The endomorphisms of $\Phi$ are given  by the following set $\{G_{\alpha}\}$ 
of {\em holomorphic} matrix-valued functions defined by \rf{aut}.
After using both the Riemann-Roch theorem, see for instance
\cite{Gun66}, and the Hawley and Schiffer results, \cite{HS68}, 
about holomorphic affine connections, we find that locally the gauge
group is parametrized, up to an overall complex number, according to
\be
\mbox{\nms in } U_{\alpha}\ :\
G_{\alpha}\ =\ \left(
\begin{array}{cc}
1 & \Omega_{\alpha} \\[2mm] 0 &  1
\end{array}
\right)
\ee
where $\{\Omega_{\alpha}\}$ are the coefficients of an abelian 
differential $\Omega = \Omega_{\alpha} dZ_{\alpha},\ 
\del_{\ovl{Z}_{\alpha}} \Omega_{\alpha} = 0$.
It is readily seen that End$(\Phi)$ is a subgroup of $SL(2,\mbf{C})$
which is not the Borel subgroup, compare with \cite{BFK91}.

Moreover, according to Gunning \cite{Gun66,Gun67a}, an endomorphism of
$\Phi$ can also be seen as a sheaf homomorphism $G:\ {\cal O}(\Phi)
\lra {\cal O}(\Phi)$, where ${\cal O}(\Phi)$ is the sheaf of germs of
holomorphic sections of $\Phi$. Here, it turns out that $\Phi$ is
represented by the first jet bundle $J_1({\cal O}(K^{-1/2}))$, where
${\cal O}(K^{-1/2})$ is the sheaf of germs of holomorphic sections
$\Psi$ of the complex line bundle $K^{-1/2}$ which are locally
represented by germs of holomorphic $(-\shalf,0)$-differentials
$\Psi\, (dZ)^{-1/2}$, $\del_{\ovl{Z}}\Psi =0$, such that
\be
\mbox{\nms in } U_{\alpha} \cap U_{\beta}\ :\
\Psi_{\alpha}\ =\ K_{\alpha\beta}^{-1/2}\, \Psi_{\beta} \ .
\ee

For a holomorphic $(1,0)$-connection, solving the glueing condition
\rf{conn} yields the following local expression
\be
\mbox{\nms in } U_{\alpha}\ :\
A_{\alpha}\ =\ \left(
\begin{array}{cc}
B_{\alpha} & \Xi_{\alpha} \\[2mm] 1 & - B_{\alpha}
\end{array}
\right)\ dZ_{\alpha} \ ,
\ee
where $\{B_{\alpha}\}$ are the coefficients of an abelian differential
and $\{\Xi_{\alpha}\}$ are well defined holomorphic objects on
$\Sigma$, see \cite{Gun67a}.  Note that Tr$A_{\alpha}=0$. However, it
is possible to parametrized locally these $\Xi$'s as $\Xi_{\alpha} =
Q_{\alpha}- B_{\alpha}^2 - \shalf P_{\alpha} -
\del_{Z_{\alpha}} B_{\alpha}$, where $\{Q_{\alpha}\}$ 
are the coefficients of a holomorphic quadratic
differential, and $\{P_{\alpha}\}$ are holomorphic functions defining
a holomorphic projective connection, \ie locally we have
$\del_{\ovl{Z}_{\alpha}} P_{\alpha} = 0$ and
\be
\mbox{\nms in } U_{\alpha} \cap U_{\beta}\ :\
P_{\beta}\ =\ K_{\alpha\beta}^{-2}\, P_{\alpha}
+ \{Z_{\alpha},Z_{\beta}\}\ ,
\ee
where $\{Z_{\alpha},Z_{\beta}\}$ denotes the Schwarzian derivative
characterizing the holomorphic projective transformations, \ie
$\{Z_{\alpha},Z_{\beta}\} = - (d^2\ln K_{\alpha\beta}/dZ_{\beta}^2
- \shalf (d \ln K_{\alpha\beta}/dZ_{\beta})^2 )$.

The gauge transformed of $A$ given by the local formula \rf{gaugetrsf}
together with the specification $\Omega_{\alpha}= - B_{\alpha}$ for
$G$ gives rise to the following unique special representative of
equivalence gauge classes of holomorphic $(1,0)$-connections, see
\cite{Gun67b},
\be
\mbox{\nms in } U_{\alpha}\ :\
\zer{A_{\alpha}} \ =\ \left(
\begin{array}{cc}
0 & - \shalf H_{\alpha} \\[2mm] 1 & 0
\end{array}
\right)\ dZ_{\alpha}\ ,
\lbl{A0}
\ee
where $H_{\alpha} = P_{\alpha} - 2Q_{\alpha}$ is thus a holomorphic
projective connection~; recall that holomorphic quadratic
differentials are the ambiguities on the determination of projective
connections. Of course the curvature $\zer{F}\,=0$. So, through this
special representative \rf{A0} one may say that holomorphic projective
connections parametrize the flat complex vector bundles of rank 2
associated to $\Phi$.

Next, following Gunning \cite{Gun67a,Gun66}, one restricts oneself to the
subsheaf of ${\cal O}(K^{-1/2})$ made of the complex analytic
solutions of the conformally covariant differential equation
\be
\Psi''_{\alpha}+ \shalf\, H_{\alpha}\Psi_{\alpha}\ =\ 0\ ,
\lbl{sol}
\ee
where the $'$ symbol stands for the holomorphic 
derivative with respect to the local coordinate $\{Z_{\alpha}\}$. This
gives germs of holomorphic sections of the bundle $K^{3/2}$.

This concludes our review on $SL(2,\mbf{C})$ complex holomorphic 
vector bundles, \cite{Gun66,Gun67a}.
But, nothing has been said yet about the action of smooth
diffeomorphisms on the base Riemann surface and the importance of
the Beltrami parametrisation of complex structures as described in
Section 3. This will be the subject of the next two points.

\subsect{The effect of the smooth change of local complex coordinates}

\indent

For a given $\mu$ \ie a given complex structure on our compact
surface, the flat holomorphic complex vector bundle $\Phi$ is
over the base Riemann surface $\Sigma_{\mu}$. 
The flat holomorphic complex vector bundle $\Phi$
will be ``pulled-back'' through the local smooth change of complex 
coordinates defined by eq.\rf{dZ} to a smooth equivalent
flat vector bundle over the Riemann surface $\Sigma\equiv\Sigma_0$. 

In more details, expressing the holomorphic coordinates $\{Z_{\alpha}\}$
in terms of the underlying complex structure $\mu$,
the sheaf ${\cal O}(K^{-1/2})$ is pulled back to the sheaf
${\cal E}(\kappa^{-1/2})$ of germs of smooth sections $\psi$ of
$\kappa^{-1/2}$ thanks to the local rescaling
\be
\mbox{\nms in } U_{\alpha}\ :\
\Psi_{\alpha}\ =\ \lambda_{\alpha}^{1/2}\, \psi_{\alpha}\ .
\ee 
This induces a local smooth change of coordinates on the fibres of
the holomorphic vector bundle~$\Phi$ 
\be
M_{\alpha}(z_{\alpha},\bz_{\alpha}) \ =\ \left(
\begin{array}{cc}
\lambda_{\alpha}^{-1/2} & 
\lambda_{\alpha}^{-1/2} \del_{\alpha} \ln \lambda_{\alpha}^{-1/2} \\[2mm]
0 & \lambda_{\alpha}^{1/2}
\end{array}
\right)\ ,
\ee
and yields a smooth vector bundle $\phi$ equivalent to $\Phi$
defined by the 1-cocycle with respect to the complex covering 
$\{(U_{\alpha},z_{\alpha})\}$,
\be
\phi_{\alpha\beta} = M^{-1}_{\alpha} \Phi_{\alpha\beta} M_{\beta}\ ,
\lbl{phi}
\ee
where, thanks to
$\lambda_{\alpha} K_{\alpha\beta} = \kappa_{\alpha\beta}\lambda_{\beta}$, 
the new holomorphic (in $z$) transition functions read 
\be
\mbox{\nms in } U_{\alpha} \cap U_{\beta}\ :\
\phi_{\alpha\beta}\ =\ \left(
\begin{array}{cc}
\kappa_{\alpha\beta}^{1/2} &
{\dps \frac{d\kappa_{\alpha\beta}^{1/2}}{dz_{\beta}}} \\[4mm]
0 & \kappa_{\alpha\beta}^{-1/2}
\end{array}
\right)\ .
\ee
Remark that $\det \phi = 1$ and $\phi = J_1({\cal E}(\kappa^{-1/2}))$
the 1-jet bundle of ${\cal E}(\kappa^{-1/2})$. 
The gauge group End$(\phi)$ is thus
defined according to the 1-cocycle $\phi$, see \rf{phi}, by 
\be
\mbox{\nms in } U_{\alpha} \cap U_{\beta}\ :\
\phi_{\alpha\beta}\, g_{\beta}\ =\ g_{\alpha} \phi_{\alpha\beta}\
\Lra\ \mbox{\nms in } U_{\alpha}\ :\ 
g_{\alpha}\ =\ M^{-1}_{\alpha} G_{\alpha} M_{\alpha}\ .
\ee
Due to the holomorphicity of $G$ and the compatibility condition
\rf{lamb}, one finds
\be
\mbox{\nms in } U_{\alpha}\ :\
g_{\alpha}\ =\ \left(
\begin{array}{cc}
1 & \omega_{\alpha} \\[2mm] 0 &  1
\end{array}
\right)\ ,\ \del_{\ovl{Z}_{\alpha}}\Omega_{\alpha}=0 \Lra
(\ovl{\del}_{\alpha} - \mu_{\alpha} \del_{\alpha})\,
\omega_{\alpha} - \omega_{\alpha}\del_{\alpha} \mu_{\alpha} = 0\ ,
\ee
where $\lambda_{\alpha}^{-1}\omega_{\alpha} = \Omega_{\alpha}$, and
$\{\omega_{\alpha}\}$ are thus coefficients of a smooth $(1,0)$-differential.

Accordingly, the flat connections on the new holomorphic vector bundle $\phi$ 
are given by 
\be
\zer{{\cal A}_{\alpha}}\ =\ M_{\alpha}^{-1} \zer{A_{\alpha}} M_{\alpha} 
+ M_{\alpha}^{-1} d M_{\alpha} \ .
\ee
Note that this is the inverse of \rf{redef} and does not correspond to
a gauge transformation~: the connections $\zer{{\cal A}_{\alpha}}$ and
$\zer{A_{\alpha}}$ are not defined on the same 
flat holomorphic vector bundle.
After the use, once more, of eq.\rf{lamb}, one finds the remarkable smooth
representative connection in more explicit form
\be
\zer{{\cal A}_{\alpha}}\ = \left(
\begin{array}{cc}
0 & - \shalf h_{\alpha}\\[2mm] 1 & 0 \end{array}
\right) dz_{\alpha} + 
\left(
\begin{array}{cc}
- \shalf \del_{\alpha} \mu_{\alpha} & - \shalf (\del_{\alpha}^2 \mu_{\alpha}
 + \mu_{\alpha} h_{\alpha}) \\[2mm]
\mu_{\alpha} & \shalf \del_{\alpha} \mu_{\alpha} \end{array}
\right) d\bz_{\alpha} \ ,
\ee
where we have set
\be
h_{\alpha}(z_{\alpha},\bz_{\alpha})\ =\ 
\lambda_{\alpha}^2 \, H_{\alpha}(Z_{\alpha}) + \{Z_{\alpha},z_{\alpha} \} \ ,
\lbl{h}
\ee
with $\{Z_{\alpha},z_{\alpha} \} = \del_{\alpha}^2 \ln \lambda_{\alpha}
- \shalf (\del_{\alpha}\ln \lambda_{\alpha})^2$ the Schwarzian
derivative of $Z$ with respect to $z$. Moreover, thanks to
eq.\rf{lamb}, one can show that
\be
(\ovl{\del}_{\alpha}-\mu_{\alpha}\del_{\alpha}- 2\,\del_{\alpha}\mu_{\alpha})
\,\{Z_{\alpha},z_{\alpha} \}\ =\ \del_{\alpha}^3\, \mu_{\alpha}\ ,
\ee
and thus, $h_{\alpha}$ fulfils the equation
\be
\del_{\ovl{Z}_{\alpha}} H_{\alpha} = 0 \Llra 
\ovl{\del}_{\alpha} h_{\alpha}= (\del_{\alpha}^3 + 2h_{\alpha}\del_{\alpha} 
+ \del_{\alpha} h_{\alpha})\, \mu_{\alpha} \lbl{compatib}\ .
\ee
The latter corresponds exactly to the vanishing of the curvature
2-form $\zer{{\cal F}\ }$ associated to $\zer{{\cal A}.}$ 
Indeed, we have locally 
$\zer{{\cal F}_{\alpha}}\ = M^{-1}_{\alpha} \zer{F}_{\alpha} M_{\alpha} = 0$. 
Moreover, due to its definition \rf{h}, the set $\{h_{\alpha}\}$ glues 
as a projective connection but it is {\em not} locally holomorphic 
with respect to the complex covering $\{(U_{\alpha},z_{\alpha})\}$.
Note that this non-holomorphic connection is local in $\mu$.

Therefore, we have obtained smooth equivalent flat holomorphic
vector bundles and the above equation \rf{compatib} expresses the compatibility
condition between the complex structure $\mu$ and the (smooth) projective 
connection $h$ defined by \rf{h}. This upshot was previously established in 
\cite{BFK91}.

Accordingly, the rescaled solution $\psi$ has to satisfy locally
\bea
\del_{\ovl{Z}_{\alpha}}\Psi_{\alpha} = 0 & \Longrightarrow & 
\ovl{\del}_{\alpha} \psi_{\alpha} = 
\mu_{\alpha} \del_{\alpha} \psi_{\alpha} - 
\shalf \psi_{\alpha} \del_{\alpha} \mu_{\alpha}\ ,
\lbl{psi} \\[2mm]
\mbox{{\nms eq.\rf{sol}}} & \Longrightarrow & 
\del^2_{\alpha} \psi_{\alpha} + \shalf h_{\alpha}
\psi_{\alpha} = 0\ .\nn
\ena
This is how the results given in \cite{BFK91} for the Virasoro
case are recovered. The present derivation makes explicit use of
the Beltrami parametrisation
and of the relationship between field variables and the components of
geometric objects via the integrating factor
$\lambda$ which is easily eliminated through eq.\rf{lamb}.

\subsect{The action of diffeomorphisms}

\indent

The Beltrami differential being explicitly introduced and involved in
a local way, we are in position for treating the action of diffeomorphisms.
The infinitesimal action of diffeomorphisms will reproduce the expected
conformal Virasoro algebra. Hence, the
``soldering'' procedure, as initiated in \cite{Pol90}, to obtain the action of
diffeomorphisms on $\Sigma$ directly from a $SL(2)$ gauge transformation on
$\Phi$ seems to remain at the level of a recipe.

The infinitesimal action of diffeomorphisms which happens to be worked out
directly from the complex coordinates $Z$ pertaining to the complex
structure $\mu$ is given locally in $U_{\alpha}$, with $s^2=0$, by,
see \rf{infgeobj},
\be
s H_{\alpha}\ =\ \gamma^{Z_{\alpha}} \del_{Z_{\alpha}} H_{\alpha},\cm
s \Psi_{\alpha}\ =\ \gamma^{Z_{\alpha}} \del_{Z_{\alpha}} \Psi_{\alpha}\ .
\lbl{sH}
\ee
In order to compute the infinitesimal action of diffeomorphisms on the
smooth projective connection $\{h_{\alpha}\}$ defined in eq.\rf{h},
one uses in sequel, on the
one hand, $sH$ given in \rf{sH} and the expression \rf{dZ} for
$\del_Z$, and on the other hand, $s\ln \lambda$
given in \rf{sz}, eq.\rf{lamb} and eq.\rf{compatib}. One finally finds
\be
s h_{\alpha}\ =\ (\del_{\alpha}^3 + 2h_{\alpha}\del_{\alpha} 
+ \del_{\alpha} h_{\alpha})\, C_{\alpha}\ ,
\lbl{sh}
\ee
which formally looks like the conformal transformation law for the
energy-momentum tensor of the Virasoro algebra.

The question, which still remains to be solved, is whether such a
$h$ might be interpreted as the energy-momentum tensor of some theory 
according to eq.\rf{compatib} which looks like the conformal Ward
identity. The latter arises from the diffeomorphism symmetry.

The following transformation law for the solutions of
eq.\rf{psi} can be added. Starting with 
$s \Psi_{\alpha}$ defined in \rf{sH} and proceeding as above
 for $h$ by using all together \rf{sz}, \rf{lamb} and \rf{psi}, one computes
\be
s\psi_{\alpha} = C_{\alpha} \del_{\alpha} \psi_{\alpha} - 
\shalf \psi_{\alpha} \del_{\alpha} C_{\alpha}\ ,
\ee
which shows that $\{\psi_{\alpha}\}$ are the coefficients of a smooth
$(-\shalf,0)$-differential. 

All these results fit perfectly together with what we expect regarding
the Virasoro algebra. There are related with the compatibilty between
complex and projective structures on a Riemann surface, upon the use
of the Beltrami parametrisation of complex structures.

\indent

\sect{On flat complex vector bundles of higher rank}

\indent

It seems very nice now that, in the Beltrami parametrisation of complex
structures, the dependence of the $SL(2,\mbf{C})$-flat connections on 
the Beltrami differential actually emerges from the above local 
process generated by a smooth change of complex coordinates.
Moreover, no constraint is imposed on the connection contrary to what is done
in the Hamiltonian reduction framework, since the use of the holomorphic 
framework insures from the very
beginning the vanishing of the curvature, see Section 2.

In view of the previous exercise, the question whether the
construction extends to higher rank vector bundle is of utmost
interest, that is, whether it works in the case of $W$--algebras. 
Indeed, the above appealing scheme (a special
representative of holomorphic connections on a flat complex vector
bundle together with the use of the Beltrami parametrisation of
complex structures on a compact Riemann surface)
fails in constructing ``generalized Beltrami differentials'' for flat
complex vector bundles of higher rank. Down below, this fact will be
illustrated in the example related to the $W_3$--algebra.

\subsect{The $SL(3,\mbf{C})$ complex vector bundle}

\indent

Let $\Phi$ be the flat complex analytic vector bundle of rank 3 defined by the
holomorphic 1-cocycle
\be
\mbox{\nms in } U_{\alpha} \cap U_{\beta}\ :\
\Phi_{\alpha\beta}\ =\ \left(
\begin{array}{ccc}
K_{\alpha\beta} & {\dps \frac{d K_{\alpha\beta}}{dZ_{\beta}}} &
{\dps \frac{d^2 K_{\alpha\beta}}{dZ_{\beta}^2}}\\[4mm]
0 & 1 & {\dps \frac{d \ln K_{\alpha\beta}}{dZ_{\beta}}}\\[4mm]
0& 0 & K_{\alpha\beta}^{-1}
\end{array}
\right) \ ,
\lbl{w3}
\ee
with $\det \Phi = 1$. After solving the cocycle condition \rf{aut}
according to \rf{w3} the endomorphisms of $\Phi$ are defined by the 
following set $\{G_{\alpha}\}$ of {\em holomorphic} matrix-valued 
functions parametrized as follows
\be
\mbox{\nms in } U_{\alpha}\ :\
G_{\alpha}\ =\ \left(
\begin{array}{ccc}
1 & \Omega^1_{\alpha} & \Theta_{\alpha} + \Omega'^2_{\alpha}
- \Omega'^1_{\alpha}\\[2mm]
0 & 1 & \Omega^2_{\alpha}\\[2mm]
0 & 0 & 1
\end{array}
\right)\ ,
\lbl{w3aut}
\ee
where $\{\Omega^i_{\alpha}\},\ i=1,2$ are the coefficients of abelian
differentials and $\{\Theta_{\alpha}\}$ are those of a holomorphic
quadratic differential.
For a holomorphic $(1,0)$-connection on that bundle, solving the
glueing condition \rf{conn} for the $SL(3,\mbf{C})$ 1-cocycle
\rf{w3} gives rise to the following local parametrization for flat
connections, in $U_{\alpha}$ :
\be
A_{\alpha}\ =\ \left(
\begin{array}{ccc}
B_{\alpha}^1 & Q_{\alpha}^1 + {B'}^2_{\alpha} - {B'}_{\alpha}^1 &
{\cal C}_{\alpha} - P'_{\alpha} + \shalf ({Q'}^2_{\alpha}-{Q'}^1_{\alpha}) -
{B''}^2_{\alpha} - 2P_{\alpha}B_{\alpha}^1\\[2mm]
1 & B_{\alpha}^2 & - 2P_{\alpha} - {B'}_{\alpha}^1 - 2{B'}^2_{\alpha} +
Q_{\alpha}^2\\[2mm]
0 & 1 & - B_{\alpha}^1 - B_{\alpha}^2
\end{array}
\right)\ dZ_{\alpha}\ ,
\ee
where $\{B^i_{\alpha}\},\ i=1,2,$ represents abelian differentials, 
$\{Q^i_{\alpha}\},\ i=1,2,$ holomorphic quadratic differentials,
$\{{\cal C}_{\alpha}\}$ are the coefficients of a holomorphic cubic
differential and $\{P_{\alpha}\}$ are those of a holomorphic 
projective connection. Note that Tr$A_{\alpha}=0$.

By choosing $\Omega^1_{\alpha}=-B^1_{\alpha},\
\Omega^2_{\alpha}=-B^1_{\alpha}-B^2_{\alpha},\ 
\Theta_{\alpha} = B^1_{\alpha}B^2_{\alpha}-Q^1_{\alpha}+Q^2_{\alpha}$
in \rf{w3aut}, and by performing the corresponding gauge
transformation on $A$ (see \rf{gaugetrsf}) one gets the special
representative for the gauge classes of holomorphic
$(1,0)$-connections
\be
\mbox{\nms in } U_{\alpha}\ :\
\zer{A_{\alpha}} \ =\ \left(
\begin{array}{ccc}
0 & 0 & {\cal C}_{\alpha} - H'_{\alpha} \\[2mm]
1 & 0 & - 2H_{\alpha} \\[2mm]
0 & 1 & 0
\end{array}
\right)\ dZ_{\alpha} \ ,
\ee
where $H_{\alpha} = P_{\alpha} + \shalf (B_{\alpha}^1B_{\alpha}^2 
- (B_{\alpha}^1+B_{\alpha}^2)^2 + Q_{\alpha}^2 - Q_{\alpha}^1)$ 
is the coefficient of a holomorphic projective connection. 
One may say that both the
holomorphic projective connections and the cubic differentials parametrize
the set of flat complex vector bundles of rank 3 associated to $\Phi$.
Notice that this special representative takes the so-called
Drinfeld-Sokolov form \cite{DS84}.

One can see that the 1-cocycle $\Phi_{\alpha\beta}$ defines the
glueing rules of $J_2({\cal O}(K^{-1}))$, the second order jet bundles
of the sheaf of germs
of holomorphic sections of the line bundle $K^{-1}$, which are then
locally represented by germs of holomorphic vector fields
$\Psi\,(dZ)^{-1}$, $\del_{\ovl{Z}}\Psi =0$, satisfying the patching rules
\be
\mbox{\nms in } U_{\alpha} \cap U_{\beta}\ :\
\Psi_{\alpha}\ =\ K^{-1}_{\alpha\beta}\, \Psi_{\beta} \ .
\ee
As considered by Gunning in \cite{Gun67a}, one can restrict the sheaf 
${\cal O}(K^{-1})$ to the subsheaf of complex analytic solutions 
of the conformally covariant differential equation
\be
\Psi'''_{\alpha}+ 2H_{\alpha}\Psi'_{\alpha} +
(H'_{\alpha} - {\cal C}_{\alpha})\Psi_{\alpha} \ =\ 0\ .
\lbl{sol'}
\ee
Note that the left-hand-side of eq.\rf{sol'} is a germ of
holomorphic sections of the bundle $K^2$.

\subsect{The Beltrami parametrisation and the action of diffeomorphisms}

\indent

As before, the flat complex vector bundle $\Phi$ is pulled back
by the smooth local change of coordinates on the fibres
\be
M_{\alpha}(z_{\alpha},\bz_{\alpha}) \ =\ \left(
\begin{array}{ccc}
\lambda_{\alpha}^{-1} &
\lambda_{\alpha}^{-1} \del_{\alpha} \ln \lambda_{\alpha}^{-1} &
(\lambda_{\alpha}^{-1} \del_{\alpha} \ln \lambda_{\alpha})^2 +
\lambda_{\alpha}^{-1} \del_{\alpha}^2 \ln \lambda_{\alpha}^{-1} \\[2mm]
0 & 1 & \del_{\alpha} \ln \lambda_{\alpha}^{-1}\\[2mm]
0 & 0 & \lambda_{\alpha}
\end{array}
\right) \ ,
\ee
induced by the local rescaling
\be
\mbox{\nms in } U_{\alpha}\ :\
\Psi_{\alpha}\ =\ \lambda_{\alpha}\, \psi_{\alpha}\ .
\ee 
Using again 
$\lambda_{\alpha} K_{\alpha\beta} = \kappa_{\alpha\beta}\lambda_{\beta}$,
yields a smooth vector bundle $\phi$ equivalent to $\Phi$
defined by the holomorphic 1-cocycle 
$\phi_{\alpha\beta} = M^{-1}_{\alpha} \Phi_{\alpha\beta} M_{\beta}$,
\be
\mbox{\nms in } U_{\alpha} \cap U_{\beta}\ :\
\phi_{\alpha\beta}(z_{\beta})\ =\ \left(
\begin{array}{ccc}
\kappa_{\alpha\beta} & {\dps \frac{d\kappa_{\alpha\beta}}{dz_{\beta}}} &
{\dps \frac{d^2\kappa_{\alpha\beta}}{dz_{\beta}^2}} \\[4mm]
0 & 1& {\dps \frac{d \ln \kappa_{\alpha\beta}}{dz_{\beta}}}
\\[4mm]
0 & 0 & \kappa_{\alpha\beta}^{-1}
\end{array}
\right) \ .
\ee
One has $\det \phi = 1$ so that
 $\phi = J_2({\cal E}(\kappa^{-1}))$.
The gauge group for $\phi$ is thus parametrized as
\be
\mbox{\nms in } U_{\alpha}\ :\
g_{\alpha}\ =\ M^{-1}_{\alpha} G_{\alpha} M_{\alpha}\Lra
g_{\alpha}\ =\ \left(
\begin{array}{ccc}
1 & \omega^1_{\alpha} & \theta_{\alpha} + 
\del_{\alpha} \omega^2_{\alpha} - \del_{\alpha}\omega^1_{\alpha}\\[2mm]
0 & 1 & \omega^2_{\alpha}\\[2mm]
0 & 0 & 1
\end{array}
\right)\ ,
\ee
where $\lambda_{\alpha}^{-1}\omega^i_{\alpha} = \Omega^i_{\alpha}$,
$i=1,2$, $\lambda_{\alpha}^{-2} \theta_{\alpha} = \Theta_{\alpha}$,
are, respectively, coefficients of smooth $(1,0)$-differentials 
and of a smooth $(2,0)$-differential with
\bea
\del_{\ovl{Z}_{\alpha}}\Omega^i_{\alpha}=0 & \Lra &
(\ovl{\del}_{\alpha} - \mu_{\alpha} \del_{\alpha})\,
\omega^i_{\alpha} - \omega^i_{\alpha}\del_{\alpha} \mu_{\alpha} = 0\
,\nn\\[2mm]
\del_{\ovl{Z}_{\alpha}}\Theta_{\alpha}=0 & \Lra &
(\ovl{\del}_{\alpha} - \mu_{\alpha} \del_{\alpha})\,
\theta_{\alpha} - 2\, \theta_{\alpha}\del_{\alpha} \mu_{\alpha} = 0\ .
\ena

Accordingly, the flat connections on the new holomorphic vector bundle $\phi$ 
are given by 
\be
\zer{{\cal A}_{\alpha}}\ =\ M_{\alpha}^{-1} \zer{A_{\alpha}} M_{\alpha} 
+ M_{\alpha}^{-1} d M_{\alpha} \ .
\ee
Note once more, this is not a gauge transformation but only a local 
change of coordinates.
More explicitly, the remarkable smooth representative of the flat
connection reads
\be
\zer{{\cal A}_{\alpha}}\ = \left(
\begin{array}{ccc}
0 & 0 & W^3_{\alpha} - \del_{\alpha} h_{\alpha} \\[2mm]
1 & 0 & -2h_{\alpha} \\[2mm]
0 & 1 & 0 \end{array}
\right) dz_{\alpha} -
\left(
\begin{array}{ccc}
\del_{\alpha} \mu_{\alpha} & \del^2_{\alpha} \mu_{\alpha} & 
\del^3_{\alpha} \mu_{\alpha} 
+ \mu_{\alpha}(\del_{\alpha} h_{\alpha} - W^3_{\alpha}) 
\\[2mm]
- \mu_{\alpha} & 0 & 2\mu_{\alpha}h_{\alpha} + \del^2_{\alpha}\mu_{\alpha}
\\[2mm]
0 & - \mu_{\alpha} & - \del_{\alpha} \mu_{\alpha} \end{array}
\right) d\bz_{\alpha} ,
\lbl{w3conn}
\ee
where $h_{\alpha}$ defined as in \rf{h} and
$W^3_{\alpha}\,\lambda^{-3}_{\alpha} = {\cal C}_{\alpha}$
fulfil respectively the following equations of vanishing curvature
\bea
\del_{\ovl{Z}_{\alpha}} H_{\alpha} = 0 &\Llra &
\ovl{\del}_{\alpha} h_{\alpha}= (\del_{\alpha}^3 + 2h_{\alpha}\del_{\alpha} 
+ \del_{\alpha} h_{\alpha})\, \mu_{\alpha} \nn\\[-2mm]
&&\lbl{0curv}\\[-2mm]
\del_{\ovl{Z}_{\alpha}} {\cal C}_{\alpha} = 0 &\Llra &
\ovl{\del}_{\alpha} W^3_{\alpha} = \mu_{\alpha} \del_{\alpha} W^3_{\alpha}
+ 3 W^3_{\alpha}\del_{\alpha} \mu_{\alpha} \nn \ .
\ena 
Notice that this special representative of flat smooth connections has
a local dependence in $\mu$.

Moreover, the rescaled solution $\psi$ has to satisfy locally
\bea
\del_{\ovl{Z}_{\alpha}}\Psi_{\alpha} = 0 & \Longrightarrow & 
\ovl{\del}_{\alpha} \psi_{\alpha} = 
\mu_{\alpha} \del_{\alpha} \psi_{\alpha} - 
\psi_{\alpha} \del_{\alpha} \mu_{\alpha}\ ,\nn\\[-2mm] 
&& \lbl{psi'}\\[-2mm]
\mbox{{\nms eq.\rf{sol'}}} & \Longrightarrow & 
(\del_{\alpha}^3 + 2h_{\alpha}\del_{\alpha} + \del_{\alpha}h_{\alpha})\, 
\psi_{\alpha} - W^3_{\alpha} \psi_{\alpha}= 0\ .\nn
\ena
At the infinitesimal level, the action of diffeomorphisms is computed
as before, and we have in addition to the variation \rf{sh},
\be
s\,\psi_{\alpha}\ =\ C_{\alpha} \del_{\alpha} \psi_{\alpha} - 
 \psi_{\alpha} \del_{\alpha} C_{\alpha}\ ,\cm
s\, W^3_{\alpha}\ =\ C_{\alpha} \del_{\alpha} W^3_{\alpha} +
3 W^3_{\alpha} \del_{\alpha} C_{\alpha}\ ,
\lbl{s3}
\ee
which shows that $\psi$ is a conformal field of weight $-1$ and that $W^3$
carries conformal weight 3. By comparison of the equations \rf{0curv},
\rf{psi'} and \rf{s3} with those obtained in \cite{GGL95}, one sees that the
former reflect the failure to obtain a ``generalized notion'' of the
Beltrami differential. Here, by a direct inspection in the (0,1) component of
the flat connection \rf{w3conn}, these generalizations can be explicitly
seen to be zero.

\indent

\sect{Conclusions}

\indent

Despite the appearance of a Beltrami differential, and the 
fact that the object $W^3$ might be the good candidate for the expected
spin 3 conformal tensor of the $W_3$--algebra, one can conclude that 
there is no
``generalized'' Beltrami differential to be read off from the 
row-column position 3-1 in the matrix of the $(0,1)$-component 
of $\zer{{\cal A}\ }$ in \rf{w3conn}. We stress 
once more that the Hamiltonian reduction scheme does not lead to the 
appearance of a Beltrami differential. However, it does seem to
provide a means  of constructing, at least formally, equations
related to the $W$--algebras.

Thus, having used the above holomorphic framework, one concludes that the
flatness condition for complex vector bundles of higher rank is too restrictive
since it enforces the vanishing of some components, \ie $\ovl{A}$. 
Rather, one should start directly with a smooth connection with
respect to a fixed complex structure on the base Riemann surface. 
In particular, this addresses the question on the way
of parametrizing the complex structures on complex vector bundles of rank
greater than two, namely, how to write down explicitly the
$(0,1)$-part of a {\em non} holomorphic connection. Some attempts in that
direction have been made, especially through the use of the so-called
Hitchin connection for Higgs bundles, \cite{BBRT91}, \cite{GJ95} and
references therein.  However, also in these references, Beltrami
differentials are ultimately identified without caring about the 
condition on their modulus to be less than one. Alternatively,
another possibility based on the use of pseudogroups will be studied 
elsewhere \cite{Laz96}. But, one should keep in
mind that the Beltrami differential originates from a smooth change of
local complex coordinates on a compact Riemann surface. The latter
gives rise to smooth equivalent flat vector bundles.

For the case of the Virasoro algebra, a geometric construction has 
been proposed in order to recover the very important Beltrami differential 
for 2-d conformal models. This construction is purely based
on the Beltrami parametrisation of complex structures 
--a point which we wish to emphasize-- through a smooth change of
local complex coordinates. The action of smooth diffeomorphisms,
the symmetry group for those models, can be easily worked out and
leads to the conformal variations computed in the physical literature.
Furthermore, this result is expected to shed light on 
the issue of the possible geometric interpretation of
$W$--algebras. The latter is highly non-trivial,
since the Virasoro algebra remains only a very particular case
and the conformal extended versions are not yet well understood.

\indent

\noindent
{\bf Acknowledgements :} The author is greatly indebted to R.~Stora 
for his constant help and for communicating to him some basic material 
based on private discussions with A.~Bilal. 
He also wishes to thank M.~Bauer and R.~Zucchini for fruitful
discussions in the early stages of this work and is deeply grateful 
to the referee for useful remarks.
Special thanks are due to the Theoretical Physics Group of the 
University of Amsterdam for its warm hospitality where this work has 
been mainly carried out under European financial support by contract
Nr.~ERBCHBICT930301.

\makeatletter
\def\@biblabel#1{#1.\hfill}

\end{document}